\documentclass[prc,twocolumn,showpacs,preprintnumbers]{revtex4}
\usepackage{bm}
\usepackage{amssymb}
\usepackage{amsmath}
\usepackage{graphicx}

\begin{document}
\title{Crystallization of Carbon Oxygen Mixtures in White Dwarf Stars}
\author{C. J. Horowitz}\email{horowit@indiana.edu} 
\author{A. S. Schneider}
\affiliation{Department of Physics and Nuclear Theory Center,
             Indiana University, Bloomington, IN 47405}
\author{D. K. Berry}
\affiliation{University Information Technology Services,
             Indiana University, Bloomington, IN 47408}

\date{\today}
\begin{abstract}
We determine the phase diagram for dense carbon/ oxygen mixtures in White Dwarf (WD) star interiors using molecular dynamics simulations involving liquid and solid phases.   Our phase diagram agrees well with predictions from Ogata et al. and Medin and Cumming and gives lower melting temperatures than Segretain et al.  Observations of WD crystallization in the globular cluster NGC 6397 by Winget et al. suggest that the melting temperature of WD cores is close to that for pure carbon.  If this is true, our phase diagram implies that the central oxygen abundance in these stars is less than about 60\%.  This constraint, along with assumptions about convection in stellar evolution models, limits the effective $S$ factor for the $^{12}$C($\alpha,\gamma$)$^{16}$O reaction to $S_{300}\leq170$ keV barns.  

\end{abstract}
\smallskip
\maketitle


Observations of cooling White Dwarf (WD) stars provide important information on the ages of stellar systems \cite{cosmochron}.  The interior of a WD is a coulomb plasma of ions and a degenerate electron gas.  As the star cools this plasma crystallizes.  Winget et al. recently observed effects from the latent heat of crystallization on the luminosity function of WDs in the globular cluster Ngc 6397 \cite{winget}.   Winget et al.'s observations constrain the melting temperature of the carbon and oxygen mixtures expected in these WD cores.  This temperature depends on the ratio of carbon to oxygen.  Therefore observations of crystallization may provide information on WD composition.

The ratio of carbon to oxygen in WD stars is very interesting.  It depends on the reaction $^{12}$C($\alpha,\gamma$)$^{16}$O.  Despite a great deal of effort, see for example \cite{c12ag}, the stellar rate for this reaction remains one of the most important unsettled rates left in Nuclear Astrophysics \cite{buchmann}.  Furthermore, the ratio of carbon to oxygen in massive stars is important for their subsequent evolution and nucleosynthesis \cite{coevolution}. Therefore, a measurement of the carbon to oxygen ratio in a WD could be very important.

To determine the C/O ratio from observations of the melting temperature one needs the phase diagram for carbon and oxygen mixtures.  Segretain et al. calculated the phase diagram assuming a local density model for the free energy of the solid \cite{segretain}.  While, Ogata et al. \cite{ogata},\cite{ichimaru} and DeWitt et al. \cite{dewitt03},\cite{dewitt96} calculated the phase diagram based on Monte Carlo or Molecular Dynamics (MD) simulation free energies for both the liquid and solid phases.  Recently Potekhin et al. have made accurate calculations of the free energy of liquid mixtures \cite{potekhin09},\cite{potekhin09b} and Medin and Cumming calculated the phase diagram for both binary mixtures such as C/O and much more complicated multicomponent mixtures \cite{mendin}.  All of these calculations are very sensitive to small errors in the free energy difference between liquid and solid phases.  Indeed Segretain et al. predict higher melting temperatures and a spindle type phase diagram while both Ogata et al. and Medin and Cumming predict lower melting temperatures and an azeotrope type phase diagram.

In this paper, we present direct two phase molecular dynamics simulations of the carbon / oxygen phase diagram to address these uncertainties.  The systematic errors of our simulations may be different from previous free energy calculations.  We discuss our formalism,  present results for the C/O phase diagram, and present possible limits on the central oxygen concentration of WDs in NGC 6397 and the effective astrophysical $S$ factor that describes the $^{12}$C($\alpha,\gamma$)$^{16}$O cross section.

We describe our two-phase MD simulation formalism.  This is very similar to what we used earlier for the freezing of rapid proton capture nucleosynthesis ash on accreting neutron stars \cite{chemsep}.   
We assume the electrons form a degenerate Fermi gas.  The ions are fully pressure ionized and interact with each other via screened Coulomb interactions.  The potential between the $i$th and $j$th ion is assumed to be
$v_{ij}(r)={Z_iZ_j e^2} {\rm e}^{-r/\lambda}/r$.
Here the ion charges are $Z_i$ and $Z_j$, $r$ is their separation and the electron screening length is $\lambda$.  For cold relativistic electrons, the Thomas Fermi screening length is $\lambda^{-1}=2\alpha^{1/2}k_F/\pi^{1/2}$ where the electron Fermi momentum $k_F$ is $k_F=(3\pi^2n_e)^{1/3}$ and $\alpha$ is the fine structure constant.  Finally the electron density $n_e$ is equal to the ion charge density, $n_e=\langle Z\rangle n$, where $n$ is the ion density and $\langle Z\rangle$ is the average charge.  Our simulations are classical and for historical reasons we have neglected the electron mass.  Including the electron mass at a White Dwarf central density of a few $10^6$ g/cm$^3$ will reduce the screening length by about 20\%.  We expect this to have only a very small effect on our computed phase diagram, however see \cite{pot1}.  Likewise quantum effects, at these densities, should also have very small effects on the phase diagram because the parameter $r_s$ describing the ratio of the ion sphere radius to the ion Bohr radius is large $r_s\approx 18000$ \cite{jones}, see also \cite{pot2}.

We now describe the initial conditions for our classical MD simulations.  It can be difficult to obtain an equilibrium crystal configuration for a large system involving a mixture of ions.  Therefore, we start with a very small system of 432 ions with random coordinates at a high temperature and cool the system a number of times by re-scaling the velocities until the system solidifies.  
Next four copies of this solid configuration were placed in the top half of a larger simulation volume along with four copies of a 432 ion liquid configuration.  The resulting system with 3456 ions was evolved in time until it fully crystalized.  Finally, four copies of this 3456 ion crystal were placed in the top half of the final simulation volume and four copies of a 3456 ion liquid configuration were placed in the bottom half.  This final system has 27648 ions and consists of a solid phase above a liquid phase.  There are two liquid-solid interfaces.  The first is near the middle of the simulation volume and the second is at the top.  This is because of the periodic boundary conditions that identify the top face with the bottom face.

The simulations can be characterized by an average Coulomb parameter $\Gamma$,
\begin{equation}
\Gamma= \frac{\langle Z^{5/3} \rangle e^2}{a_e T}\, .
\label{gammamix}
\end{equation} 
Here $\langle Z^{5/3} \rangle$ is an average over the ion charges, $T$ is the temperature, and the electron sphere radius $a_e$ is $a_e=(3/4\pi n_e)^{1/3}$.

All of our simulations are run for the same electron density of $n_e=5.026\times 10^{-4}$ fm$^{-3}$.    Since the pressure is dominated by the electronic contribution, constant electron density corresponds, approximately, to constant pressure.  Ignoring quantum effects, the density can be scaled to other values by also changing the temperature $T$ so that the value of $\Gamma$, see Eq. \ref{gammamix}, remains the same.    

We have performed six simulations with parameters as indicated in Table \ref{tableone}.  We evolve the system in time using the simple velocity Verlet algorithm \cite{verlet} with a time step $\Delta t=25$ fm/c for the pure carbon simulation and 100 fm/c for the five carbon/ oxygen mixture simulations.  We use periodic boundary conditions.  Our simulation volume is large enough so that the box length $L$ is much larger than the electron screening length $\lambda$. The ratio of the force on two ions separated by a distance $L/2$, compared to the force on two ions separated by the ion sphere radius $a=(3/4\pi n)^{1/3}$ is $F(L/2)/F(a)\approx 10^{-5}$.  

\begin{table}
\caption{Computer Simulations with 27648 ions.  The carbon number fraction for the whole system is $x_c$, the total simulation time is $t$, and the final temperature is $T$.  The carbon number fraction of the solid phase is $x_c^s$, while $x_c^l$ is the carbon fraction of the liquid phase.}
\begin{tabular}{llllll}
Run & 
$x_c$ & $t$ (fm/c) & $T$ (MeV) & $x_c^s$ & $x_c^l$\\
 
 c1 & 
 1& $5\times 10^7$& 0.02050(3)  & 1& 1\\

 c90 &
 0.90 & $8\times 10^8$ & 0.0195(1) & 0.906(11) & 0.900(8) \\
  
 c824 & 
 0.824 & $1\times 10^9$ & 0.0197(1)  & 0.834(5) & 0.819(5)\\ 
 
 c75 & 
 0.75 & $2\times 10^9$ & 0.0193(1) & 0.727(5) & 0.766(5)\\
 
 c50 & 
 0.50 & $2\times 10^9$ & 0.0217(1) & 0.459(5) & 0.525(5) \\
 
 c25 & 
 0.25 & $2\times 10^9$ & 0.0256(1) & 0.192(4) & 0.284(4) \\

\end{tabular} 
\label{tableone}
\end{table}


We first describe the pure carbon simulation, run c1 in Table \ref{tableone}.  We start by evolving the 27648 ion system at constant temperature for a time of a few million fm/c.  During this time, we carefully adjust the temperature, by rescaling the velocities, so that about half of the system is solid and half is liquid.  Then we evolve the system at constant energy for 50 million fm/c.  
Finally, as long as the potential is independent of momentum, the expectation value of the kinetic energy per particle is $3T/2$.  Therefore we estimate the melting temperature from the kinetic energy.  This yields a melting $\Gamma$ value of
\begin{equation}
\Gamma_m=178.4 \pm 0.2\, ,
\label{gammam}
\end{equation}
see Table \ref{tableone}.  This value is slightly larger than the $\Gamma_m=175$ expected for the One Component Plasma (OCP) \cite{pot1}\cite{dewitt} and consistent with the value $\Gamma_m\approx 178.2$ predicted by Eq. (4) in ref. \cite{yukawa} for a Yukawa system with our value of the parameter $\kappa=1/(n^{1/3}\lambda)=0.542$ (and assuming $\Gamma_m=175$ for the OCP). 

Our two-phase simulation method is subject to systematic errors from finite size effects and from non-equilibrium effects due to the finite run time.  Finite size effects could be important because the simulation volume contains two liquid-solid interfaces so that an ion may be relatively close to one of the interfaces.  Simulations in ref. \cite{chemsep} with 3456 ions yielded a melting $\Gamma$ that is only slightly smaller $\Gamma_m=176.1 \pm 0.7$.   We conclude that finite size effects are relatively small for our 27648 ion single component system.

We search for non-equilibrium effects by evaluating the temperature at times of 10, 20, 30, 40, and 50 million fm/c.  We find very little time dependence.  Therefore, we expect non-equilibrium effects to be small for our single component system.  

Our simulations for carbon / oxygen mixtures are performed in a similar way.  We start from an initial configuration that is half liquid and half solid.  The initial number fraction of carbon $x_c$ in the solid phase is equal to that in the liquid phase.  The system is evolved in time, first at constant temperature and then at constant energy.  The carbon and oxygen ions are free to diffuse across the liquid-solid interfaces so that the number fraction of carbon in the liquid $x_c^l$ can become different from the number fraction in the solid $x_c^s$.

\begin{figure}[ht]
\begin{center}
\includegraphics[width=3.5in,angle=0,clip=true] {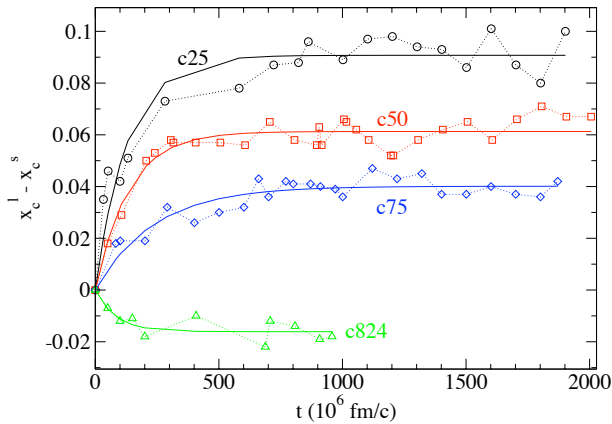}
\caption{(Color on line) Number fraction of carbon in the liquid phase minus the number fraction of carbon in the solid phase versus simulation time for the simulations of Table \ref{tableone}.}
\label{Fig2}
\end{center}
\end{figure}

We now present results for the phase diagram of carbon and oxygen mixtures.  We measure the composition of the liquid $x_c^l$ and solid $x_c^s$ as follows.  We divide the simulation volume into fifteen regions equally spaced in $z$ coordinate and calculate the average $\langle Z \rangle$ for each region.  Adjacent regions where $\langle Z \rangle$ changes from below to above average are assumed to represent liquid-solid interfaces and their composition is discarded.  The remaining 11 regions are used to calculate the average composition of the liquid $x_c^l$ and solid $x_c^s$.  Figure \ref{Fig2} shows the difference $x_c^l-x_c^s$ versus simulation time.  Note that there are relatively large statistical errors.
The solid curves in Fig. \ref{Fig2} are least squares fits of the form $x_c^l-x_c^s=a[1-{\rm exp}(-b\, t)]$ with $a$ and $b$ constants.

We see that non-equilibrium effects can be significantly larger for mixtures, because it can take a long time for impurities to diffuse.  We note that the solid is enriched in oxygen, compared to the average composition, for runs c75, c50, and c25, while for run c824 the solid is enriched in carbon, compared to the average composition.  Simple estimates of diffusion times suggest that the concentration should have equilibrated by the relatively long simulation time of $2\times 10^9$ fm/c.  Runs c75, c50, and c25 were performed on special purpose MDGRAPE-2 hardware \cite{mdgrape} and took approximately six months of computer time each.      

\begin{figure}[ht]
\begin{center}
\includegraphics[width=3.5in,angle=0,clip=true] {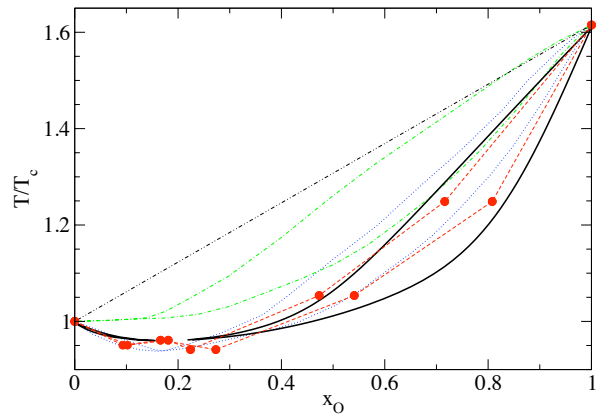}
\caption{(Color on line) Melting temperature $T$ of carbon / oxygen mixtures over the melting temperature $T_c$ of pure carbon, versus oxygen number fraction $x_o=1-x_c$.  Simulation results of Table \ref{tableone} are plotted as filled red circles connected by dashed lines.  The $x_o$ in the liquid and in the solid are shown as two separate lines.  Also shown are the phase diagram results of Medin and Cumming \cite{mendin} as solid black lines, the Ogata et al. results \cite{ogata} as dotted blue lines and the Segretain et al. results \cite{segretain} as dot-dashed green lines.  Finally the black dot-dot-dashed line corresponds to $\Gamma=178.4$ in Eq. \ref{gammamix}.}
\label{Fig3}
\end{center}
\end{figure}

We average the liquid and solid compositions over the final approximately 400 million fm/c of simulation time to determine the carbon-oxygen phase diagram.  These results are listed in Table \ref{tableone} and plotted in Fig. \ref{Fig3} as filled red circles.  The upper curve gives the composition of the liquid that is in equilibrium with a solid of composition given by the lower curve.  Note that the run c1 is plotted twice in Fig. \ref{Fig3}, first at $x_o=0$ (pure carbon) and then rescaled to $x_o=1$ (pure oxygen).   We find that the melting temperature of carbon oxygen mixtures is considerably below the constant $\Gamma_m=178$ prediction, Eqs. \ref{gammamix},\ref{gammam}.  This is plotted as a dot-dot-dashed line in Fig. \ref{Fig3}.  Our melting temperatures are also below the results of Segretain et al. \cite{segretain}.  We speculate that this could be because of small errors in Segretain et al's density functional calculations of the solid free energy.

Our results agree qualitatively with Ogata et al. \cite{ogata}, and, in general, agree well with Medin and Cumming \cite{mendin}.  Both of these calculations are based on Monte Carlo or MD simulation free energies for the liquid and solid phases.  Although the overall agreement with Medin and Cumming is good, there is a tendency for our simulations to predict smaller differences in composition between the liquid and solid phases $x_o^l-x_o^s$.  This could be because of finite size effects in our simulations.  In equilibrium, there is a composition gradient, as a function of position, across the liquid-solid interface.  Therefore, if one probes the composition of the liquid and solid in positions that are too close to the interface, one will naturally get smaller differences between $x_o^l$ and $x_o^s$.    Alternatively, $x_o^l-x_o^s$ could be sensitive to small errors in Medin and Cummings' free energies.  Overall, given the agreement between our results and those of Ogata et al. and Medin and Cumming, we conclude that the carbon / oxygen phase diagram is largely known and that it is of azeotrope, instead of spindle, form.

We now discuss implications of our carbon oxygen phase diagram on White Dwarf (WD) star crystallization, limits on the oxygen fraction of WDs, and possible limits on the $^{12}$C($\alpha,\gamma$) reaction rate.  Winget et al. observe the luminosity function (number of WD with a given luminosity versus luminosity) for the globular star cluster NGC 6397 \cite{winget}.  They find a peak in the luminosity function that they attribute to crystallization, and they claim that the location of the peak is sensitive to the crystallization temperature of the WD core.    
  
Winget et al.'s observations agree well with theoretical luminosity functions for 0.5-0.535 $M_\odot$ WDs with pure carbon cores.  The observations disagree with a theoretical luminosity function assuming a WD core of 50\% carbon and 50\% oxygen by mass (or $x_o=0.43$ by number).   This luminosity function fixed the melting temperature with Eqs. \ref{gammamix},\ref{gammam}, \cite{winget_private}, for which $T=1.26T_c$ at $x_o=0.43$, see the dot-dot-dashed line in Fig. \ref{Fig3}.

For simplicity, we assume theoretical luminosity functions can be characterized by the melting temperature of the core.  The data clearly favor $T$ near $T_c$ and strongly disfavor $T=1.26T_c$.  If, in the future, one could set a limit of, for example, half this difference, $T<1.13 T_c$ by comparing theoretical luminosity functions to observations, then we can place limits on core oxygen concentrations.  These limits follow from the form of our phase diagram in Fig. \ref{Fig3}.  The melting temperature is close to $T_c$ for $x_o < 0.5$ and then rises rapidly with increasing oxygen concentration.  If one had a constraint of $T<1.13T_c$ then our phase diagram implies
\begin{equation}
x_o<0.57
\end{equation}
for the oxygen concentration by number or $X_o<0.64$ for the oxygen concentration by mass.  {\it We conclude that constraining the melting temperature of WD cores to be close to that for pure carbon constrains the oxygen concentration to be of order 60\% or less.} 

Salaris et al. find that the oxygen concentration in WD cores depends on the cross section for the $^{12}$C($\alpha,\gamma$)$^{16}$O reaction, that can described by the astrophysical $S$ factor, and on the treatment of convection in stellar evolution models \cite{salaris}.  With their treatment of convection and an effective $S$ factor, at an energy of 300 keV, of $S_{300}=240\ {\rm keV\, barns}$, Salaris et al. find $X_o=0.79$ for the oxygen concentration in the core of a 0.54 $M_\odot$ WD.  This oxygen concentration would be ruled out if $T<1.13T_c$.  Alternatively, Salaris et al. find $X_o=0.57$ for a 0.6 $M_\odot$ WD if they assume a smaller value $S_{300}=170 \ {\rm keV\, barns}$.  We expect the central oxygen abundance of a 0.54 $M_\odot$ star to be slightly larger than that for a 0.6 $M_\odot$ star and close to our limit of $X_o<0.64$.  We conclude that assuming $T<1.13T_c$, along with the Salaris et al. assumptions for convection, implies a limit on the effective $S$ factor of order
\begin{equation}
S_{300}\leq 170\ {\rm keV\, barns}.
\end{equation}
This limit is consistent with the recent experimental determination by Buchmann and Barnes of $S_{300}=145$ keV b, with an error of 25\% to 35\%,  \cite{buchmann}.  It is also consistent with the results of Tur, Heger and Austin who evaluate nucleosynthesis yields by varying both the triple-alpha and $^{12}$C($\alpha,\gamma$) rates \cite{tur}.  Their best fit value is $S_{300}=174$ keV b, with perhaps significant error.

In the future, WD luminosity functions should be calculated using our carbon oxygen phase diagram.  Possible limits on crystallization temperatures should be deduced from observations including careful consideration of systematic errors.  Finally, we will study the role of a small neon abundance on the phase diagram and WD crystallization using three component MD simulations, see for example \cite{segretain1996}. 



We thank E. Brown, A. Chugunov, A. Cumming, H. DeWitt, Z. Medin, and D. Winget for helpful discussions.  This research was supported in part by DOE grant DE-FG02-87ER40365, by Shared University Research grants from IBM, Inc. to Indiana University, and by the National Science Foundation through TeraGrid resources provided by National Institute for Computational Sciences, and Texas Advanced Computing Center under grant TG-AST090112.

\vfill\eject


\begin{thebibliography}{99} 
\bibitem{cosmochron} G. Fontaine, P. Brassard, P. Bergeron, PASP {\bf 113}, 409 (2001).
\bibitem{winget} D. E. Winget et al., ApJ. {\bf 693}, L6 (2009).
\bibitem{c12ag}R. Kunz et al., ApJ {\bf 567}, 643 (2002).
\bibitem{buchmann} L. R. Buchmann, C. A. Barnes, Nuc. Phys. {\bf A777}, 254 (2006).
\bibitem{coevolution} M. F. El Eid, B. S. Meyer, L. S. The, ApJ {\bf 611}, 452 (2004).
\bibitem{segretain} L. Segretain, G. Chabrier, A\& A, {\bf 271}, L13 (1993).
\bibitem{ogata} S. Ogata, H. Iyetomi, S. Ichimaru, PRE {\bf 48}, 1344 (1993).
\bibitem{ichimaru} S. Ichimaru, H. Iyetomi, S. Ogata, ApJ {\bf 334}, L17 (1988).
\bibitem{dewitt03}H. DeWitt, W. Slattery, Contrib. Plasma Phys. {\bf 43}, 279 (2003).
\bibitem{dewitt96}H. DeWitt, W. Slattery, G. Chabrier, Physica B {\bf 228}, 21 (1996).
\bibitem{potekhin09} A. Y. Potekhin, G. Chabrier, F. J. Rogers, PRE. {\bf 79}, 016411 (2009).
\bibitem{potekhin09b} A. Y. Potekhin, G. Chabrier, A. I. Chugunov, H. E. DeWitt, F. J. Rogers, arXiv:0909.3990.
\bibitem{mendin}Z. Medin, A. Cumming, submitted to PRE and private communication 2009.
\bibitem{chemsep} C. J. Horowitz, D. K. Berry, and E. F. Brown, Phys. Rev. E {\bf 75}, 066101 (2007).
\bibitem{pot1} A. Y. Potekhin, G. Chabrier, Phys. Rev. E {\bf 62}, 8554 (2000).
\bibitem{jones} M. D. Jones, D. M. Ceperley, PRL. {\bf 76}, 4572 (1996).
\bibitem{pot2} A. Y. Potekhin, Doctor of Science thesis (in Russian) http://www.ioffe.ru/astro/DTA/palex/disser.pdf and to be published in Contrib. Plasma Phys.
\bibitem{verlet} L. Verlet, Phys. Rev. {\bf 159}, 98 (1967).
\bibitem{mdgrape} T. Narumi, R. Susukita, T. Ebisuzaki, G. McNiven, and B. Elmegreen, Molecular Simulation 21, 401 (1999).
\bibitem{dewitt} H. DeWitt, W. Slattery, D. Baiko, D. Yakovlev, Contrib. Plasma Phys. {\bf 41}, 251 (2001).
\bibitem{yukawa} O. Vaulina, S. Khrapak, G. Morfil, Phys. Rev. E {\bf 66}, 016404 (2002).
\bibitem{winget_private} D. Winget, private communication, 2010.
\bibitem{salaris} M. Salaris et al., ApJ. {\bf 486}, 413 (1997).
\bibitem{tur} C. Tur, A. Heger, S. M. Austin, ApJ. {\bf 671}, 821 (2007).
\bibitem{segretain1996} L. Segretain, A\&A. {\bf 310}, 485 (1996).

\end{thebibliography}
\end{document}